# THE IMPACT OF ARTIFICIAL INTELLIGENCE ON ENTERPRISE SOFTWARE USER ROLES

AI and Enterprise Software User Roles

An Empirical Case Study of Role Transformation on the SAP Business Technology Platform

Isabel Unger

SAP SE, isabel.unger@sap.com

Elizangela Valarini

SAP SE, elizangela.valarini@sap.com

Martin Schrepp

SAP SE, martin.schrepp@sap.com

Nina Hollender

SAP SE, nina.hollender@sap.com

Gabriela Rocha

SAP SE, gabriela.guasti.diniz.rocha@sap.com

Erik Bertram

SAP SE, Fresenius Hochschule Heidelberg, erik.bertram@sap.com

Artificial Intelligence (AI) is rapidly reshaping the nature of work in software development, transforming user roles, workflows, and collaboration patterns across enterprise platforms. This qualitative study investigates how AI alters professional responsibilities within the context of SAP's Business Technology Platform (BTP), combining expert interviews (n=20) and a participatory workshop (n=24). The results reveal substantial shifts in day-to-day tasks and roles in the development domain, characterized by increasing automation of operational tasks, expanding human–AI collaboration, and growing reliance on agentic AI systems. The study further identifies significant implications for existing user-role frameworks, such as the BTP User Type Matrix, which requires adaptation as the workforce is undergoing significant role specific changes. Collectively, these findings highlight a workforce landscape in transition and underscore the need for revised role taxonomies, new governance and oversight functions, and updated design approaches for AI-native enterprise software systems.

## 1 INTRODUCTION

The rapid emergence of Artificial Intelligence (AI) tools—particularly in the software development sector—has initiated a profound transformation of the workforce in job profiles related to software development. Tools based on Artificial Intelligence, such as coding assistants and agentic systems, see an exponential growth on the market and are increasingly integrated into daily workflows, reshaping how software is designed and developed.

Simultaneously, organizations are reinforcing these developments by investing in AI and driving AI adoption. Using AI is seen as an indispensable strategic advantage. 88% of companies report active investment in AI agents [1]. Moreover, 74% of companies indicate that their AI use cases are already generating measurable business value.

Overall, the evolution of technologies and corresponding offerings from software providers is reshaping the nature of work for employees interacting with these systems. These shifts impact workflows, job structures, and work content, particularly within the software development domain [2,3]. Anticipating future workforce transformations, 92% of companies expect that managing AI agents will become a critical skill within the next five years [1]. Internal research findings from Anthropic claims that AI is reshaping job profiles to an unprecedented degree [4]. These dynamics highlight a growing need for in-depth research into the future evolvement of roles within the development domain. It is of course quite difficult to make clear predictions about future developments in such a fast-developing domain like AI. And there are clearly different expectations between providers of AI tools and AI infrastructure and their customers.

In this study, we attempt to uncover expectations of experienced professionals concerning the impact of AI on their current responsibilities and professional environments, as well as how AI is shaping their roles. Section 2 presents an overview of the current research landscape concerning the impact of AI on developer roles. Section 3 gives a short overview about the product environment underlying the case study, introducing the SAP Business Technology Platform (BTP) and the Framework BTP User Type Matrix. Section 4 presents the used methodology, methods, and dataset. In Section 5, we discuss the results of the triangulated data and subsequently, in Section 6, we discuss the findings and limitations of this work.

## 2 LITERATURE AND RELATED WORK

The integration of Large Language Models (LLMs) and agentic AI tools into software engineering has initiated a paradigm shift in how code is authored, validated, and maintained. This transformation encompasses significant changes to the development lifecycle and a fundamental evolution of the programmer's professional role [5,3,6].

Some authors assume that this introduction of AI capabilities into traditional workflows will also massively change the role of human users [7,8]. People who developed or controlled complex systems or planned large technical infrastructures will hand over parts of their tasks to an AI system. Generative AI fundamentally changes how users interact with digital applications: deterministic workflows in a standardized user interface shift to the formulation of an “intent”. AI acts as an assistant or agent system and contributes to the result in a non-deterministic manner, which users evaluate and iteratively refine. Thus, their role changes transitioning towards a co-worker of an AI system [9]. In addition, completely new roles or job opportunities will emerge.

The transition to an AI-supported working mode however requires that the technology is accepted by users. Compared to classical software applications, new criteria for user experience must be considered to capture human needs [10,8].

To reach the desired gain in productivity, users must hand over some of their tasks to an AI and in many cases the AI is a black box from the point of view of the user. Thus, the right amount of trust in the AI system [11] is important to avoid users spending as much time checking the systems' decisions and actions as they would need if they would do the tasks completely on their own. On the other hand, users need to cultivate some skepticism to be able to control the AI and stay in the driver's seat. However, to develop trust in AI systems, these need to provide some human-centered AI qualities [7], for example transparency [12], explainability [13] as well as the ability of the users to stay in control [14,15].

Very recent studies have shown that the introduction of AI technologies such as “vibe coding”, is shifting development tasks through acceleration and automation, and experienced developers reject the “blind trust” associated with the term in favor of rigorous oversight [16]. While the developer's daily activities are increasingly automated, ensuring software quality remains a human-centric activity requiring established engineering principles.

With the increasing induction of AI into the workforce, potential lacks in code quality and security arise [17, 18]. AI struggles with maintainability, structure, secure patterns and long-term quality. In addition, AI-generated code, partially even created by the enhancement of coding democratization pose new challenges in the areas of security and governance [19]. Regulatory frameworks such as the EU AI Act or the ISO standards impose requirements on the development and deployment of AI systems to ensure standards for AI systems, e.g. in the fields of security, governance, ethics [20, 21]. These increased and new needs pose demands on roles in the development domain.

For software developers or administrators of complex technological landscapes, AI offers various benefits and opportunities to increase their productivity. Especially the transformer architectures and the large language models (LLMs) built on them allowed fundamentally new possibilities [22]. Tools like GitHub Copilot and ChatGPT make AI-supported development accessible to a wider developer community. AI support covers the entire software development lifecycle and is used, for example, for code completion and generation, the explanation of unknown code sections, the creation of unit tests, debugging, the automatic generation of documentation, or assistance with architecture planning [23]. A particularly dynamic field is the so-called Automatic Program Repair (APR), where models like GPT-4 can identify faulty lines of code and directly integrate appropriate corrections into the source code [24].

Empirical studies show productivity gains by AI-driven development. AI assistants can help developers to complete tasks up to 55% faster, significantly reducing the time spent on “grunt work” [16]. This acceleration is primarily achieved through the automation of repetitive and scaffolding tasks, such as generating boilerplate code, writing unit tests, and drafting documentation [25].

AI is changing the development landscape, shifting from implementation to architectural oversight. Additionally, AI is transforming the technical maintenance landscape and quality assurance toward the automation of routine software engineering activities. However, AI is not only impacting responsibilities of professionals in the software development domain, it is also impacting team dynamics and learning loops substantially. As a sparring partner, developers are increasingly turning to AI automatically reducing the interaction with the human peers [16].

Examining AI adoption across the entire software development lifecycle, Alenezi and Akour [2025] demonstrate that AI use is concentrated primarily in the development and testing phases, as reflected by the frequency of tool utilization. Across these stages, developers perceive improvements in efficiency, code quality, and team collaboration enabled by AI [26]. According to Ulfsness et al. [2024], working with existing code—Including refactoring or simplifying code, code review, translating code—was the most mentioned use case among development-related roles. Other “GenAI activities” centered around general assistance, learning, creating non-technical boilerplate, presentations and mail writing, demonstrating, that AI is applied in a wide range of activities.

Perceived effects of AI use vary by task complexity (simple versus advanced tasks). Ulfsness et al. [2024] found different benefits and challenges of AI usage. Benefits were identified for aspects such as more time for learning, efficiency gains and pair programming. Challenges were seen in areas such as Input cleaning necessities, lack of tool integrations, limited context ability and the need for competence in effective prompt engineering.

Huang et al. [2025] provide complementary insights into the usage of AI Agents in software development practices. Their findings indicate that experienced developers see the value that AI Agents can contribute to the coding process, but they maintain their decision-making authority in software design and implementation. Experienced developers express positive attitudes towards AI Agents in development as they are confident to compensate for a potential lack of accuracy or quality of AI generated outputs.

Several studies underline the impact of code generating models on developer productivity. Despite initial productivity drains due to processes such as trainings, environment configurations, multiple studies show an increased productivity for developers working with AI [25, 27].

Rising efficiency gains sparked debate over the future of software development roles, given AI's increasing ability to perform routine coding tasks. Nonetheless, multiple analyses conclude that AI cannot replace human developers [28, 29]. Complementing these perspectives, empirical evidence suggests that the productivity impact of AI may be more limited outside controlled laboratory environments. A study involving nearly five thousand developers across three companies—conducted during regular, real-world work—provides strong ecological validity for this conclusion [30]. Nevertheless, many studies highlight significant time savings and enhanced efficiency when AI supports coding activities [31].

Beyond the notable shifts in software development practices driven by the integration of Artificial Intelligence, other studies reveal how this transformation is reshaping the roles in software development.

A growing body of literature indicates the role of the software developer to be transitioning from a "writer of code" to a "controller and reviewer" [4, 15, 5]. Experienced developers now strategically oversee AI behavior, maintaining their agency through rigorous planning and active supervision. This relationship is often bimodal: Acceleration and exploration modes. In the acceleration mode, the developer acts as the "driver", using AI as a super-charged autocomplete to finish well-understood microtasks without breaking their cognitive flow. In contrast to this, in the exploration mode, the developer lets the AI "drive" to help navigate unfamiliar APIs, brainstorm algorithms, or get unstuck on novel problems, essentially using the tool as a "sparring partner" for rubber-ducking [5]. Sauvola et al. define this as "human-in-the-loop" [32].

Consequently, prompt engineering and communication have emerged as critical new skills [33, 34]. Success with AI tools heavily relies on the developer's ability to provide clear context, explicit instructions, and detailed specifications. Developers must also exercise rigorous oversight to mitigate risks like "automation bias" or the "user-synthesizer gap", where a user might over-rely on a tool that produces superficially correct but logically flawed code [35].

Findings from Anthropic [2026] indicate that developers increasingly transition toward full-stack roles as capabilities increase, especially in more junior roles. For example, developers who previously focused on backend development only were now coding UIs. Also, as AI increases capabilities, the cross-role collaboration is impacted, too. For instance, backend developers were able to generate complex UIs with AI support, illustrating how AI can blur traditional boundaries between roles.

Beyond productivity, AI adoption also appears to influence collaboration patterns and the overall functioning of development teams. With AI systems increasingly functioning as quasi-team members, their integration reshapes team interactions, communication structures, and the experiential dimension of work, partly similarly to human teammates [9, 36]. Still, it should be noted that these artificial relationships can cause emotional fatigue if an AI is not capable of emotionally responding to a human [37].

Potential decrease in knowledge sharing and communication may negatively affect team collaboration and performance [3]. AI further influences workflows, collaboration structures, and learning dynamics in agile development teams and complex architectural setups. While this increases individual efficiency, it may reduce the collective knowledge sharing and inter-team interactions that characterize healthy agile environments. Despite these risks, the shift toward agentic workflows is generally viewed as the future of the industry, making the development process more enjoyable for those who can effectively steer the technology.

**Analysis of Big Cloud Platforms**

The impact of generative AI use cases is among the highest across all industries, underscoring its transformative potential for organizations worldwide [2]. In response, major cloud providers are revising their strategic directions to align with an agent centric paradigm. AI has become a central strategic priority for major software vendors, cloud hyperscalers, and enterprise platform providers. AI agents capable of collaborating, interacting with software systems, and autonomously completing end to end workflows are emerging as the new standard in future software systems.

A review of strategic developments at companies demonstrates several overarching trends in the software development domain. These include democratization of AI capabilities, design of agents for workflow automation, agent oversight, advances in security, consideration of ethical principles, and new approaches to agent orchestration [38, 39, 40]. Technical advancements further accelerate this shift as agent architectures can execute entire workflows autonomously and interacting with enterprise software systems [41].

Microsoft Azure is developing towards agentic AI with Azure AI Foundry serving as a central pillar for developing and scaling agentic AI systems [42]. The systems are capable of complex task execution with long-term memory and continuous oversight over AI actions through the Foundry Control Pane. The platform enables no-code / low-code development and therefore democratizes AI development for professional developers and business users. Developers work primarily with Azure OpenAI and Azure AI Foundry, while business users interact through Copilot Studio and Power Apps.

Salesforce is evolving from a traditional CRM system into an orchestration platform for autonomous AI systems [43], centered on the Agentforce Platform, which enables the development, oversight, and execution of autonomous AI agents independently execute business processes. Developers and data scientists interact via Data Cloud and agent-building environment, while business users engage through Slack. In parallel, IBM's watsonx supports developers in building, scaling, and integrating AI into existing workflows [46].

SAP is similarly evolving toward an agent-centric enterprise platform, with Joule serving as an orchestration layer for intent-driven interaction and autonomous process execution across enterprise systems. Joule Studio will help transform how developers work and accelerate the way agents, applications, extensions and workflows are built. Intent based coding support to create and extend agents, apps and applications.

These developments demonstrate that AI is going to be deeply embedded in software development workflows, whereas AI agents are expected to be taking over complete workflows. In addition, collaboration patterns might change as business users and developers collaborate differently if development is democratized among these roles. Platforms are becoming increasingly agent-centric, expanding the range of tasks that can be taken over by AI. The insight into tools and technical capabilities and strategical visions of different leading companies in the industry demonstrate that there is a big impact on development roles through AI. The following section presents an overview of the current research landscape concerning how these technological advancements are impacting roles in the software development domain.

## 3 BTP USER TYPE MATRIX: A CASE STUDY

This section presents a case study of the SAP Business Technology Platform, examining the impact of Artificial Intelligence on key user roles and tasks within a cloud platform context [44]. Both the platform and its user base are undergoing significant transformation.

SAP is restructuring user interaction through Joule Work and the autonomous suite, enabling end-to-end processes in an intent-based workspace across applications. Joule Studio provides an AI-based environment for developing agents and workflows grounded in enterprise data, supporting autonomous execution across SAP and external systems, and combining low-/no-code with pro-code, using prompt-based design and AI-assisted code generation, and includes capabilities for integration, orchestration and governance [44, 45].

To investigate the impact of AI on software development roles, we conducted a qualitative case study approach focusing on the evolution of key user roles—namely, the primary users of BTP products and tools. As BTP users operate across multiple stages of the development lifecycle, analyzing changes in their tasks and activities provides insights into how roles evolve in response to increasing AI integration. In this context, the BTP User Type Matrix serves as a relevant lens for understanding the transformation of roles in the software development domain.

### 3.1 Framework: The User Type Matrix

The BTP User Type Matrix is a framework designed to support alignment among stakeholders involved in the development lifecycle of BTP applications and services, including UX designers, developers, architects, cloud administrators, data scientists, and product managers. It provides a structured representation of high-level User Types, organized along two axes: task categories (y-axis) and interaction paradigms ranging from no-code to pro-code (x-axis) (see Figure 1 below).

By providing a structured representation of high-level User Types across task categories and interaction paradigms, the matrix fosters a shared understanding of users, their needs, and their journeys across interconnected applications, rather than within isolated products. This enables more consistent, user-centered design and development decisions, reduces fragmentation across teams, and supports scalable design practices in complex platform ecosystems. Target stakeholders therefore are all roles involved in product development, especially UX-Designers, Developers and Product Managers.

Usually, in UX-Design the concept of personas is most frequently used to describe users. Personas are research-based, yet fictional archetypes to guide product decisions by providing actionable information on users' goals, tasks and pain points [46, 47]. When looking at a suite of applications developed by different teams, each team may have their own set of detailed personas. To support alignment across teams about users, and to avoid mental overload through to much detail, we apply User Types in addition to personas. They are higher level; research grounded abstractions that cluster personas into cross product categories.

User Types are close to the concept of roles [48]. We chose the term "User Type" over the term "role" to make a clear distinction to role-based access control (RBAC), which is a technical term used in the enterprise software context to manage system access [49].

The term role can be defined as the set of behaviors expected from a person occupying a particular position within a group [50]. Accordingly, an individual may take on multiple roles simultaneously [48]. This understanding aligns with our research regarding User Types: Real users typically represent more than just one User Type. In professional settings, the term job role is frequently used in a similar manner. When referring to roles within this paper, the term "role" refers to these job-oriented User Types

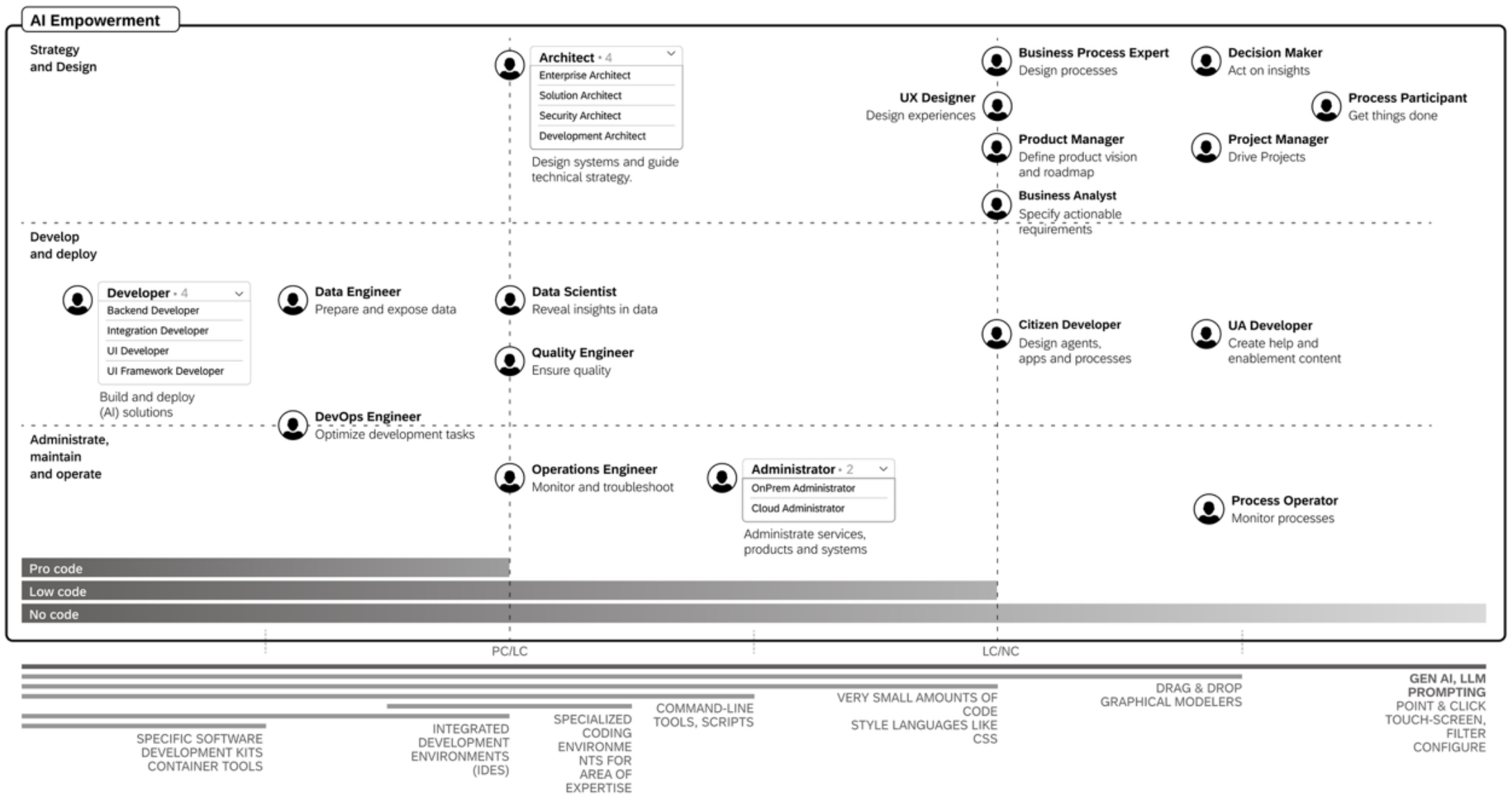

Figure 1: Visualization of the BTP User Type Matrix[1]

### 3.2 Objective and Research Questions

The objective of this research is to gain a comprehensive understanding of how AI is reshaping the nature of work for professionals in the software development domain. To this objective, the study aims to:

(1) Delineate the current state of AI adoption by characterizing its day-to-day usage across various roles in the context of software development—in this case study particularly across BTP. This involves an examination of AI tool adoption, its integration into existing workflows, and its subsequent impact on work outputs.

(2) Understand the impact of AI on the landscape of software development roles in the near future. This analysis explores the evolution of professional tasks and roles within an AI-augmented context, with a particular focus on the implications of agentic AI capabilities for low-code and no-code development practices.

Based on the objectives of this study, our research questions are as follows:

R1: What patterns of AI tool usage can be observed among SAP BTP users today, and how do these patterns vary across user roles?

R2: How will AI impact the entire user experience of SAP BTP and change user roles, skills, and tasks?

The insights derived from this study are expected to be generalizable beyond the SAP BTP context, as user roles and underlying task objectives are largely consistent across cloud platforms (e.g., SAP BTP, Salesforce). Consequently, the findings contribute to a broader understanding of how professional roles in the software development domain evolve in response to the increasing adoption of AI-native technologies and provide a foundation for informing the design of future software systems.

---

[1] Designed by Jasmin, SAP SE

## 4 METHODS AND SAMPLE

Using a qualitative approach, we collected and analyzed empirical data to map the changes in the User Types of the Business Technology Platform and their corresponding tasks. We combined semi-structured expert interviews and workshops, complemented by a small quantitative survey embedded within the workshop to provide additional context on evolving roles and tasks associated with AI.

The qualitative methods enabled access to intangible data, such as thoughts, expectations, and feelings associated with AI usage [51]. Additionally, the qualitative approach allowed us to explore how AI is currently being used and how it will be used in the future, as well as identify existing activities, practices and perceived changes. The table below gives an overview of the methods, key objectives, and target groups among SAP and non-SAP professionals in the software development domain.

Table 1: Overview of the Research Approach

| Method | Expert Interviews | Workshop |
|---|---|---|
| Sample | 20 | 24 |
| Role | Machine Learning Engineer<br>AI Scientist<br>Development Architect<br>AI Developer<br>Development Manager<br>UX-Designer<br>Data Science Expert | Developers<br>Product Owners<br>Development Architects<br>Technical- and Business Consultants |
| Objective | Explore the current AI usage, changes on the activities due to AI adoption, as well as future expectations. | Understand and describe user roles and the associated tasks Explore anticipated future tasks per role across the development lifecycle and assess perceived levels of AI involvement. |

### 4.1 Expert Interviews

To address the research questions, this study employed semi-structured guideline-based Expert interviews [52, 53]. Participants were selected from a pool of SAP BTP users and professionals who have significant AI expertise or regularly engage with AI in their development roles. The selection was based on the premise that their experiences might possibly point to future usage scenarios. To mitigate potential biases associated with future-oriented speculation, the study prioritized individuals with high AI context awareness. This approach is supported by literature suggesting that greater AI literacy fosters more realistic estimations of its future impact on one's profession [54]. The more aware people are of their future work, the less afraid they are to interact with AI [55]. People who thought less about their role and future AI interaction might be less realistic in these regards.

The interviews were conducted between September and November 2025 and explored both current AI usage and future expectations. The interview guideline consisted of two main sections. The first section addressed participants' present AI usage, including attitudes, tools, and perceived work impact, to gather data for the first research question (RQ1). The second section explored anticipated changes in tasks, AI adoption, and the emergence of new roles, in alignment with the second research question (RQ2).

Interviews ranged from 29 to 56 minutes in duration and were transcribed verbatim. The data was subsequently analyzed by a thematic analysis [56]. This method combined deductive main categories derived from the research objectives with subcategories that were developed through both deductive and inductive coding of the interview transcripts.

### 4.2 Workshop

Building on the expert interviews, an onsite workshop of 45 minutes was conducted with participants in development-oriented roles who have prior experiences with or a stated interest in Artificial Intelligence.

To better understand workshop participants' user roles and associated tasks, a small quantitative survey (n =19) was conducted. The survey was used as a complementary instrument embedded within the workshop to support the interpretation of the results and to generate an unbiased understanding of participants' work practices and roles. It comprised two parts: first, participants described their work practices and roles without predefined categories; second, they selected predefined development roles (based on the BTP User Types in Figure 1) matching their own role and those of collaborators and added any missing roles.

The workshop participants represented five different job roles and were organized into corresponding groups: Developers, architects, product owners, technical consultants and business consultants. The objective of the workshop was to explore both current and anticipated future tasks for each role across the development lifecycle, and to assess the perceived level of AI involvement in those tasks.

To achieve this, the workshop was split into two phases. In the first phase, participants used sticky notes to map their current professional tasks onto five predefined development lifecycle stages: "Research & Plan", "Design", "Develop", "Test & Rollout", "Maintain & Operate" (x-axis). They also rated the degree of AI involvement for each task on a continuum ranging from entirely "human-controlled" to fully "AI-controlled" (y-axis). In the second phase, participants repeated this exercise for tasks they anticipated performing in the future. In this design, the workshop draws on generative methods that use structured activities to capture participants' experiences and expectations. The dual-phase design facilitated a comparative analysis of present versus expected AI integration across different roles and development stages.

The workshop data was analyzed using a mixed-methods approach. Qualitative analysis involved the examination of visual patterns and thematic clustering of sticky notes on the posters. This was supplemented by a quantitative analysis of the frequency and distribution of notes to measure task volume and changes in perceived AI involvement.

## 5 RESULTS

The following section presents the overall results derived from the study, answering the two main research questions (RQ1) addressing the patterns of AI tool usage today and the (RQ2) expected impact of AI on user roles, skills and tasks in future.

### 5.1 General Insights

Regarding the prospective use of AI and its implications for development activities and roles, the interviews indicate three main domains of change. (1) AI adoption is expected to reconfigure core tasks through increased automation, expanded collaboration with AI agents, and a reduced emphasis on manual coding. (2) Collaboration patterns are likely to evolve, encompassing both human–human coordination and human–agent interaction, leading to shifts in responsibilities and team structures. (3) Organizations will face increasing demands related to security and regulatory compliance, alongside changes in UX practices, collectively reshaping role boundaries and fostering the emergence of new roles.

When it comes to the evolution of roles, the research revealed two overarching trends:

- Role broadening and overlap: AI expands induvial capabilities and responsibilities across roles and blurs traditional role boundaries.
- Emergence of new specialized roles: AI introduces new risks, responsibilities, and tasks, leading to the development of new specialized roles.

In this regard, AI will create new roles across both data- and development-oriented domains while simultaneously transforming existing roles through new skill requirements. A particularly significant impact is expected in human–agent collaboration, with interviewees across all roles describing AI agents not merely as tools but as active collaborators. These agents are perceived as a distinct category of "software users" capable of autonomously executing development-related tasks. An analysis of the evolution of professional roles in the software development domain further highlights emerging requirements in both data- and development-related fields. In addition, participants emphasized the need for overarching roles to address system-level challenges arising from AI integration, including areas such as AI governance and orchestration.

### 5.2 AI Integration into Work Processes

When analyzing the integration of AI into workflows today, certain tasks can be observed across all roles. These primarily involve project management, learning and research tasks. Furthermore, role-specific tasks with AI support were identified for each role. Findings are shown in Table 2 below.

Table 2: Usage Patterns: Main Current Tasks per Role with AI Support

| General Taks | Architects | Developers | Product Owners | Consultants | Data Scientist |
|---|---|---|---|---|---|
| Research | Coding | Coding | Coding | Coding | Coding |
| Learning | Decision making | Code analysis | Testing | Solution | Debugging |
| Business-related tasks | Architecture design | Testing | Database management | investigation | Research e.g. model |
| Translations | suggestion | Documentation | Incident management | Architecture | identification |
| Operational tasks | Governance & | Code refactoring | Agent building | Suggestion | |
| Project definition | Security | | Onboarding | | |
| | Documentation | | | | |

Overall, AI appears to be used mainly in more streamlined, operational tasks of less complexity, such as coding, research and test generation. There was a high overlap in AI-tools used among participants. These included LLM tools, such as Chat GPT and Perplexity, code generation tools, especially GitHub Copilot and Claude Code, and AI features in various design tools.

For future tasks, the results reveal a clear tendency towards more complex tasks being executed by AI, bringing the human more into a strategical and oversight-focused role. Table 3 lists tasks that are expected to be accomplished with AI in the future. The task list does not constitute an exhaustive representation of the future responsibilities associated with the respective roles; rather, it should be interpreted as an indication of potential future developments and emerging trends.

Table 3: Usage Patterns: Main Future Tasks per Role with AI Support

| General Taks | Architects | Developers | Product Owners | Consultants | Data Scientist |
|---|---|---|---|---|---|
| Research | Coding | Coding | Technology implementation | Coding | Coding |
| Learning | Decision making | Feasibility analysis | System management | Deterministic tasks | Business understanding |
| Business-related tasks | Architecture design suggestion | Translation of apps | Process improvement | Testing | Data preprocessing |
| Translations | Cost optimization | Ideation | Financial steering | Incident analysis | |
| Operational tasks | Governance & Security | Error analysis | | Metric observation | |
| Project definition | Documentation | Bug fixing | | Information gathering | |
| | | End-to-end troubleshooting | | User/ access management | |

The findings clearly indicate a transition of developer activities towards oversight, prompting and quality assurance, with AI acting as a collaborator across the software development lifecycle, while still requiring human oversight. The findings suggest that developer expertise remains very central, particularly of senior developer roles, in the future. Development architect positions, commonly filled by more experienced employees remain important. As for developers, human-completed tasks remain central, especially when it comes to decision making upon AI generated suggestions, such as context aware architectural recommendations. The findings show a separation of tasks between highly AI assisted tasks, and human oversight and decision making on the other hand. Data-related roles experience significant impacts from AI, particularly with the emergence of new roles in AI development, generating and testing models, etc. The connection between development and data roles grows even closer with the integration of AI into their work. The survey data from the workshop indicates that these roles are often close, as they are often fulfilled by one and the same person. For Product Owners as well as Consultants, AI will play an integral role across both operational and strategic and monitoring activities. Identified tasks suggest that the Product Owner is in practice significantly more integrated with development activities than commonly assumed. The findings clearly indicate that, within these roles, no distinct tasks emerged that could not be supported or complemented by AI.

On the skill-level, the findings indicate that AI is broadening skills of individual roles not only within their actual job role, but also across roles. Designers can bring their designs to code, blurring the line into frontend development. At the same time, there were some concerns about skill degradation due to AI, which went hand in hand with the partial disappearance of junior roles, while participants emphasized the need for highly skilled workers, especially in development roles to ensure code quality. Despite the concern about skill degradation, participants also mention enhanced learning opportunities through AI.

Several impacts on the roles of participants can be concluded from the findings. Interviews suggest that participants expect role distinctions to diminish as AI expands competencies and responsibilities. Developers who were previously specialized—such as backend engineers—anticipate being able to perform a broader range of tasks with AI support, approximating a full-stack profile. Survey results collected from workshop participants, align with this trend, showing that respondents frequently selected multiple development roles or identified themselves as full-stack developers. Similar patterns emerged with business- and project-oriented roles, where combinations such as Product Owner, Citizen Developer, Project Manager, and Business Process Expert were common. Overall, the findings point to a broadening of role definitions and a reduced degree of specialization.

However, the studies also identified a growing need for new specialized tasks and therefore potentially new roles across both development and data science domains. In software development, emerging roles might include AI Architect and AI

Developer. In data-related fields, roles such as AI Scientist, Machine Learning Engineer, AI Administrator, AI Architect and Machine Learning Operations specialist might be gaining importance. Additionally, there is an upcoming need for cross-functional roles addressing overarching concerns such as AI governance, security, and agent orchestration.

Key needs identified include:

1. AI Agent Orchestration: A role focused on managing and coordinating multiple AI agents.
2. AI Governance and Control: Functions dedicated to ensuring the security, compliance, and ethical performance of AI systems.
3. System Review and Maintenance: A persistent human task focused on the oversight and validation of AI-driven outputs and processes.

Overall, the findings suggest that role definitions remain fluid and are not yet clearly established, reflecting the ongoing transformation driven by AI integration. Therefore, it is not yet clear which roles will come up as new roles, and which will be reflected within tasks and responsibilities of existing roles.

Table 4 provides an overview of role evolvements detected within the study, by providing information on the current state and future changes per role.

Table 4: Overview of AI Adoption: Current Use and Future Implementation among User Roles

| Role | Developers | Development Architects | Product Owners | Data Scientists | Consultants |
|---|---|---|---|---|---|
| General statements | - AI is increasingly embedded across nearly all stages of the software development lifecycle acting as companion rather than replacement<br>- The developer role is shifting from hands-on coding toward oversight and quality assurance.<br>- Human oversight remains essential, particularly in supervising AI-generated outputs and actions. | - Decision-making remains predominantly human-driven despite increasing automation.<br>- AI serves as an augmenting tool, providing advanced, context-aware recommendations.<br>- This support extends to complex tasks such as architectural design.<br>- Final judgment continues to reside with human actors. | - AI usage is increasing significantly, particularly in the automation of operational tasks.<br>- Strategic and planning activities, as well as handling of complex processes are expected to become highly AI-supported.<br>- As a result, product owners are likely to shift their focus toward long-term planning and performance monitoring. | - AI is expected to automate many operational tasks, particularly in coding and data handling.<br>- The role is becoming more central in managing and evaluating AI systems. | - AI is increasingly supporting more streamlined and operational tasks.<br>- Almost only AI supported tasks have been anticipated by consultants for the future |
| Current stage | - AI currently supports planning and research activities.<br>- It contributes to learning and architectural decision-making.<br>- It is applied in project management tasks.<br>- It assists with development tasks such as coding, refactoring, and testing. | - AI supports architects in project management, decision-making, and research.<br>- Assists with software development tasks, including coding, testing, and documentation.<br>- This support is expected to continue, with increased emphasis on governance and security in the future. | - Current AI applications support project management.<br>- AI assists with research activities.<br>- It provides support for coding and testing.<br>- AI contributes to application and incident management.<br>- It aids in onboarding processes. | - Responsibilities increasingly include model oversight, data governance, and the evaluation of machine learning (ML) and large language models (LLMs), with a stronger focus on generating measurable business outcomes. | - AI is used for easy and time-consuming tasks such as research, mail writing, documentation and test creation.<br>- AI is only partially consulted for specific tasks. |
| Future changes | - Future AI use is expected to remain largely consistent.<br>- AI involvement in more complex tasks is expected to increase.<br>- Developer roles are likely to converge over time. | - AI is expected to automate routine activities and handle more complex and autonomous tasks.<br>- Increased AI involvement enhances decision-making processes.<br>- Role is likely to evolve toward more strategic responsibilities.<br>- Human judgment will be combined with AI-generated insights. | - Future AI applications will have stronger focus on technology implementation, system management, and process optimization.<br>- AI will play an integral role across both operational and strategic activities.<br>- AI will improve efficiency, reduce risks, and enable real-time decision-making. | - New specialized roles are emerging within the data domain, including AI Scientist and ML Engineer. | - AI is heavily supporting all tasks of consultants in future.<br>- Support is anticipated for tasks such as research, technical architecture drafts, coding, user management, meeting scheduling.<br>- Trends suggest an increase in strategic work. |

## 6 DISCUSSION

Subsuming the results, participants expressed mixed feelings regarding AI, especially regarding efficiency, reflecting ongoing debates in the literature about the mixed findings about efficiency associated with AI adoption [30, 31].

Analyzing todays AI integration in work processes, it appears to be used mainly in more streamlined, operational tasks of less complexity, such as coding, research and test generation, across the whole development lifecycle, which is in line with findings from other studies [23, 26]. In the future, AI is envisioned to support more complex tasks and task bundles with agentic potential, such as AI-assisted troubleshooting or code review with direct recommendations for improvement. Overall, there was a tendency towards tasks in the "Maintain and Operate" phase of the development lifecycle.

The findings show clear differences between mainly business oriented (Product Owner and Consultant) and mainly technically oriented roles (Developers, Architects, Data Scientists) and their expected future use of AI. Development roles identified several tasks they do not foresee collaborating on with AI, whereas business roles reported none. This suggests notable variation in anticipated AI integration across roles, potentially explained by limited abilities of individual participants to envision AI usage which requires further research.

Apart from a few domain-specific differences, there was substantial overlap in the use of development-related tools by participants. This potentially provides further indication for an ongoing democratization of tasks—particularly coding—enabled by AI [38, 42, 43].

Job role developments are characterized by two main trends in the era of AI: A movement toward greater specialization due to new risks and tasks and, simultaneously, a broadening of responsibilities through AIs amplification of capabilities. The findings as well as literature and job postings analysis confirm the lack of a unified role taxonomy [57, 58, 59]. This tension presents challenges for the labor market and for anticipating how user roles and work practices will evolve in the future.

### Methodological Reflections

Selecting participants with prior connection to and experience in AI placed a risk for confirmation bias and halo effects [60, 61]. Nevertheless, given the difficulty of obtaining reliable insights into future work, the participants are considered a reasonable approximation to potential future development-related roles. Prior research also indicates that a comprehensive understanding of AI might enable participants to give more accurate estimations of AI's future impact on work [54].
The qualitative sample comprised 44 participants results can therefore just be interpreted as tendencies or trends. Considering these constraints, an additional research cycle would have strengthened the robustness and validity of the conclusions.

## 7 IMPLICATIONS FROM THIS STUDY

### Blurring Domains and Changing User Roles on the BTP User Type Matrix

The analysis indicates that AI induces significant changes on the different levels of the BTP User Type Matrix. As skill boundaries increasingly blur, exemplified by business roles engaging in coding, the traditional categorization into pro-code, low-code, and no-code roles becomes less meaningful. Instead, a continuum emerges along the x-axis of the matrix, ranging from no-code to pro-code roles. As AI enables roles to perform tasks that originally lay within the scope of other roles (such as designers generating coded UI elements), a similar continuum appears along the y-axis.

The findings also show the need for new tasks, potentially leading to new roles within the data and development domain. Simultaneously, the research suggests highly specialized roles to converge and consequently better be represented as broader, overarching role profiles—for example, the shift toward full-stack-like development roles.

These developments indicate that a new way of user visualization may be required, as the conceptual axes of the current matrix become increasingly blurred. Two options seem to come up: Defining highly specialized users or broad role profiles with more detailed skill and task profiles. Future research and design practice should tackle the question of how to represent users and AI agents in a framework in this continuously evolving user landscape.

**Practical Implications from the study**

The study reveals a work landscape already substantially reshaped by AI, with participants expressing heterogeneous perceptions of AI use. The findings suggest that organizations revisiting their role frameworks should adopt a dual approach: responsibilities across roles are broadening, leading to the emergence of higher-level job groups, while simultaneously requiring dedicated AI-focused roles in areas such as security and other specialized domains.

At the skill level, T-shaped competence models appear promising for defining future capability needs, distinguishing broad foundational expertise and deep domain knowledge in one area [62]. The research findings can serve as an insight to understand how different user groups imagine using AI. This can be considered when designing software systems.

## 8 CONCLUSION AND OUTLOOK

Overall, the study results mirror the dynamic evolution of the industry due to AI, especially in the software development domain. It aimed to make overall statements on how role frameworks in the development sector could evolve with AI, taking the case study of the SAP BTP user roles. Therefore, two qualitative studies have been conducted to understand dynamics and role evolutions.

The study indicates that user roles and tasks of roles are dynamically evolving, impacting team- and cross-role collaboration patterns. Roles in development evolve towards specialization on the one hand and broadening of responsibilities on the other hand. This should be researched further.

Attention needs to be placed on collaboration impacts, such as changing hand-offs and interaction patterns, as well as social cohesion and emotional safety in teams. Furthermore, the integration of AI pose organizations another critical challenge to the traditional knowledge transfer. As AI systems automate an increasing number of tasks and reshape team collaboration, the person-to-person transfer mechanisms are disrupted. Consequently, it remains an open and critical question of how organizations will sustain the transfer of implicit knowledge in future.

Given the ongoing fluidity in job roles, ongoing research needs to be done to understand changing roles and responsibilities.